\documentclass[12pt,preprint]{aastex}

\def\beq{\begin{equation}}
\def\eeq{\end{equation}}
\def\v{\upsilon}

\shorttitle{SSD Can Smoothly Match ADAF}
\shortauthors{Lu et al.}

\begin{document}

\title{Shakura-Sunyaev Disk Can Smoothly Match
Advection-Dominated Accretion Flow}

\author{Ju-Fu Lu, Yi-Qing Lin, and Wei-Min Gu}

\affil{Department of Physics, Xiamen University, Xiamen 361005, China; 
lujf@xmu.edu.cn}

\begin{abstract}
We use the standard Runge-Kutta method to solve the set of basic equations 
describing black hole accretion flows composed of two-temperature plasma. 
We do not invoke any extra energy transport mechanism such as thermal 
conduction and do not specify any ad hoc outer boundary condition for the 
advection-dominated accretion flow (ADAF) solution. We find that in the 
case of high viscosity and non-zero radiative cooling, the ADAF solution 
can have an asymptotic approach to the Shakura-Sunyaev disk (SSD) solution, 
and the SSD-ADAF transition radius is close to the central black hole. 
Our results further prove the mechanism of thermal instability-triggered 
SSD-ADAF transition suggested previously by Takeuchi \& Mineshige and 
Gu \& Lu.
\end{abstract}


\section{Introduction}

The most famous model of accretion disks is the Shakura-Sunyaev disk (SSD; 
Shakura \& Sunyaev 1973). Since SSD was constructed exactly 30 years ago, 
the most important breakthrough in the field of accretion disk theory has 
been the proposal of advection-dominated accretion flow (ADAF; Narayan \& 
Yi 1994; Abramowicz et al. 1995). SSD and ADAF appear to be adequate to 
describe the outer and the inner region of black hole accretion flows, 
respectively, and a phenomenological SSD+ADAF model has been quite 
successfully applied to black hole X-ray binaries and galactic nuclei 
(see Narayan, Mahadevan, \& Quataert 1998 for a review). In this model, 
however, a smooth transition from an outer SSD to an inner ADAF was only 
assumed, but not proved. From the physical point of view, the question 
remains whether (and how) an SSD can really match an ADAF.

There have been basically three classes of answers to this question. 
Dullemond \& Turolla (1998) and Molteni, Gerardi, \& Valenza (2001) gave 
negative answers, arguing that a smooth transition from an SSD to an ADAF 
was not possible. Whereas their conclusions were under some certain 
conditions: Dullemond \& Turolla (1998) considered only the low-viscosity 
case (with the viscosity parameter $\alpha \sim 0.1$); and Molteni 
et al. (2001) referred only to the plain ADAF, i.e. that with zero cooling. 
The second class of answers, on the other hand, is positive. A number of 
authors showed that the SSD-ADAF transition was realizable if an extra 
heat flux caused by thermal conduction was invoked either in the radial 
direction (Honma 1996; Manmoto \& Kato 2000; Gracia et al. 2003), or in the 
vertical direction (Meyer \& Meyer-Hofmeister 1994; Meyer, Liu, \& 
Meyer-Hofmeister 2000). The cost of this class of answers is, in our 
opinion, the involvement of an additional mechanism of energy transport, 
and in particular, the introduction of a new unknown parameter $\alpha_T$ 
to measure thermal conduction (Manmoto \& Kato 2000). The third answer 
was due to Takeuchi \& Mineshige (1998) and Gu \& Lu (2000), who suggested 
that the thermal instability of a radiation pressure-supported SSD could 
trigger the flow to jump from the SSD state to the ADAF state. 
This answer is also a positive one, but without involving any extra 
mechanism of energy transport. 

In this Letter, we present our answer to 
the question of SSD-ADAF transition. We demonstrate that such a transition 
in a smooth way is possible for flows with large values of $\alpha$
(different from the case of Dullemond \& Turolla 1998) and non-zero 
radiative cooling (different from the case of Molteni et al. 2001). We 
do not involve any extra energy transport mechanism such as conduction, 
and this is different from the above mentioned second class of 
answers, and is similar to the third answer. We discuss in some detail 
the relation between our results here and those of Takeuchi \& Mineshige 
(1998) and Gu \& Lu (2000).

\section{Equations}

The dynamical equations for steady state axisymmetric accretion flows we 
consider here are usual in the literature (e.g. Narayan, Kato \& Honma 1997). 
That is, the continuity, radial momentum, vertical equilibrium, and 
angular momentum equations read
\beq
\dot M = -4\pi R H \rho \v \ , 
\eeq
\beq
\v \frac{d\v}{dR} = \Omega^2 R - \Omega_K^2 R - \frac{1}{\rho}\frac{dp}{dR}
\ , 
\eeq
\beq
H = \frac{c_s}{\Omega_K} \ ,
\eeq
\beq
\frac{d\Omega}{dR} = \frac{\v \Omega_K (\Omega R^2 - j)}{\alpha R^2 c_s^2}
\ ,
\eeq
where $\dot M$ is the constant mass accretion rate; $R$ is the radius; 
$H$ is the half-thickness of the flow; $\rho$ is the density of the 
accreted gas; $\v$ is the radial velocity; $\Omega$ is the angular velocity; 
$\Omega_K$ is the Keplerian angular velocity, and 
$\Omega_K^2 = GM / (R-R_g)^2 R$ in the well known Paczy\'{n}ski \& Wiita (1980) 
potential, with $M$ being the mass of the central black hole, and $R_g$ being 
the gravitational radius; $p$ is the pressure; $c_s$ is the isothermal sound 
speed of the gas, defined as $c_s^2 = p / \rho$ ; and $j$ is an integration 
constant that represents the specific angular momentum accreted by the 
black hole. 

As for the energy equation, we employ the form given by Narayan \& Yi (1995), 
which is for flows composed of two-temperature plasma with bremsstrahlung 
and synchrotron radiation and Comptonization, i.e. a relatively complete and 
complex case of black hole accretion flows:
\beq
Q_{vis} = Q_{adv} + Q_{Cou} \ ,
\eeq
\beq
Q_{Cou} = Q_{rad} \ .
\eeq
These two equations are for the energy balance of the ions and of the 
electrons, respectively.
$Q_{vis}$ and $Q_{adv}$ are the rate of viscous heating that is primarily
given to the ions and the rate of advective cooling by the ions, and are
expressed by, e.g. equations (5) and (6) of Gu \& Lu (2000), respectively.
$Q_{Cou}$ is the rate of energy transfer from the ions to the electrons 
through Coulomb collisions, and is expressed by equation (3.3) of
Narayan \& Yi (1995).  $Q_{rad}$ is the rate of radiative cooling of 
electrons, and is calculated using a bridging formula which is valid
in both optically thick and optically thin regimes,
\beq
Q_{rad} = 8 \sigma T_e^4 \left ( \frac{3\tau}{2} + \sqrt {3}
+ \frac{8 \sigma T_e^4}{Q_{br} + Q_{sy} + Q_{br,c} + Q_{sy,c}}
\right ) ^{-1} \ ,
\eeq
where $T_e$ is the electron temperature; $\tau$ is
the total (election scattering plus absorption) optical 
depth, $\tau = \tau_{es} + \tau_{abs} = (0.34 cm^2 g^{-1})\rho H 
+ (Q_{br} + Q_{sy} + Q_{br,c} + Q_{sy,c}) / 8 \sigma T_e^4$ ; and 
$Q_{br}$, $Q_{sy}$, $Q_{br,c}$, and $Q_{sy,c}$ are the cooling rates 
of bremsstrahlung radiation, synchrotron radiation, Comptonization of 
bremsstrahlung radiation, and Comptonization of synchrotron radiation, 
and are explicitly expressed by equations (3.4), (3.18), (3.23), and (3.24) 
of Narayan \& Yi (1995), respectively. 

Finally, the equation of state is needed to close the system of equations,
\beq
p = p_g + p_r + p_m \ ,
\eeq
where $p_g = \Re \rho (T_i+T_e)$ is the gas pressure, $T_i$ is the ion
temperature; $p_r = Q_{rad} (\tau + 2/\sqrt{3})/4c$ is the radiation
pressure, and $p_m = B^2 / 8\pi$ is the magnetic pressure, $B$ is the 
magnetic field, and for simplicity it is usually assumed that
$\beta_m \equiv p_m/(p_g+p_m) = {\rm const}$.

Note that the dynamical equations (1)-(4) are valid for both geometrically 
thin and thick flows (Narayan et al. 1997), and equation (7) is a convenient
interpolation between the optically thin and thick limits. When the flow
is extremely 
optically thick, equation (7) gives $Q_{rad} = 16\sigma T_e^4/3\tau$, 
which is the appropriate black body limit; whereas in the optically thin 
limit it gives $Q_{rad} = Q_{br} + Q_{sy} + Q_{br,c} + Q_{sy,c}$. Thus we 
expect that the above set of equations can be used to verify the 
possible transition from an (optically thick, geometrically thin) SSD to 
an (optically thin, geometrically thick) ADAF.

\section{Numerical Solutions}

There are nine equations (equations [1]-[6], [8] plus the definition 
$p = \rho c_s^2$ and the expression of $\tau$) for nine unknown variables 
$\v$, $\Omega$, $c_s$, $H$, $\rho$, $p$, $T_i$, $T_e$ and $\tau$ as 
functions of $R$, with $M$, $\dot M$, $\alpha$, $j$, $\beta_m$ and the
adiabatic index $\gamma$ being constant flow parameters. 
We use the standard Runge-Kutta method to solve the set of three 
differential equations (2), (4), and (5) for three unknowns $\v$, $\Omega$, 
and $c_s$, and then obtain other variables from the remaining algebraic 
equations. We integrate the three differential equations from the sonic 
point $R_s$ (where the radial velocity is equal to the sound speed) both 
inward and outward. The derivatives $d\v /dR$, $d\Omega /dR$, and 
$dc_s/dR$ at $R_s$, which are needed to start the integration, are 
evaluated by applying the l'H\^{o}pital rule. The inward, supersonic part 
of the solution extends to the inner boundary of the flow, i.e. to a 
radius $R_{\rm in}$ where the no-torque condition $d\Omega /dR = 0$ 
(i.e. $\Omega R^2 = j$) is satisfied. More important for our purpose here 
is the outward, subsonic part of the solution. It should be stressed that 
we do not specify any ad hoc outer boundary conditions. We just observe 
how the outward solution evolves with increasing $R$. On the other hand, we
obtain a standard SSD solution that is calculated from a set of purely
algebraic equations (e.g. Frank, King, \& Raine 2002). The given flow
parameters $M$, $\dot M$, $\alpha$ and $j$ in the SSD solution are
exactly the same as those in the above solution obtained with the
Runge-Kutta method. We watch if and where the two solutions can smoothly
match with each other.

Figure 1 provides an example of global solution of accretion flow,
i.e. the flow quantities as functions 
of $R$. The solid line represents the solution of the nine equations in
$\S$ 2, with given parameters $\alpha = 0.7$,
$\dot m = 0.01$ ($\dot m \equiv \dot M / \dot M_{\rm Edd}$, 
with $\dot M_{\rm Edd}$ being the Eddington accretion rate), 
$j = 0.742 (cR_g)$, $\gamma = 1.5$, $\beta_m = 0.5$, and $R_s = 2.95R_g$; 
and the dashed line represents the SSD solution with the same parameters
$\alpha$, $\dot m$, and $j$. Note that $R_s$ is not another free parameter,
it is the eigenvalue of the problem, and is self-consistently determined
when the constant flow parameters are given.
Figure 1(a) is for the radial velocity $\v$ and the sound speed $c_s$.
It is seen that the solid line solution is transonic, with the sonic point
being marked by a filled square; while the SSD solution is subsonic
everywhere, it alone cannot describe the transonic nature of black hole 
accretion.
Figure 1(b) shows the angular momentum $l$~($ = \Omega R^2$). The SSD solution
follows the Keplerian distribution $l_K$~($ = \Omega _K R^2$), and
the solid line solution is sub-Keplerian.
Figure 1(c) draws the flow's relative thickness $H/R$, which
is $\sim 0.4$ (geometrically thick) for small $R$, decreases as $R$ increases,
and reaches to $\sim 0.005$ (geometrically thin) of the SSD solution.
Figure 1(d) is for the optical depth $\tau$, again with increasing $R$ , 
the flow becomes from being optically thin ($\tau \ll 1$) to being optically 
thick ($\tau \gg 1$, the SSD solution).
Figure 1(e) is for the ion temperature $T_i$ and the 
electron temperature $T_e$. The solid line solution has $T_i \gg T_e$; and
as $R$ increases, the two temperatures drop down and become identical
(the SSD solution).
In Figure 1(f) one sees that the advection factor 
$Q_{adv}/Q_{vis} = (Q_{vis}-Q_{rad})/Q_{vis}$ is $\sim 1$ (advection-dominated)
for small $R$, and decreases dramatically with increasing $R$, and reaches
nearly zero finally (radiative cooling-dominated, the SSD solution). From
these figures we conclude that the solid line solution is an ADAF solution
as it has properties of transonic radial motion, sub-Keplerian rotating,
and being geometrically thick, optically thin, very hot, and of course,
advection-dominated; and that this ADAF solution does match with the
(dashed line) SSD solution, forming together a global solution.
If the transition radius $R_{tr}$ is defined so that 
$\tau = 1$ there, then $R_{tr}\approx 12R_g$ in the solution of Figure 1. 

\section{Discussion}

We have shown that a smooth SSD-ADAF transition is realizable for black hole 
accretion flows with high viscosity ($\alpha = 0.7$ in Figure 1) and 
non-zero radiative cooling. Our argument is simple and naive. The equations 
we solve and the numerical method we use are usual. We do not introduce any 
extra energy transport mechanism such as thermal conduction. Perhaps the 
only tool somewhat special here is the bridging formula (7) expressing 
the radiative cooling $Q_{rad}$, which we need to join the optically 
thick regime to the optically thin regime of the flow. 

Our Figure 1 looks very similar to Figure 1 of Manmoto \& Kato (2000),
a representative paper of the second class of answers to the question of
SSD-ADAF transition as mentioned in Introduction. However, the similar
results are obtained in different ways: (1) As mentioned already,
Manmoto \& Kato (2000) invoked radial thermal conduction and, in particular,
introduced a new unknown parameter $\alpha_T$ to measure this extra heat
transport mechanism; while we do not. (2) They used the relaxation method
to solve the differential equations, and we adopt the Runge-Kutta method.
In principle, the solution obtained should not be related to the numerical
method, but different methods suit solving different problems.
In order to have a solution for the subsonic flow between the sonic point
and the outer boundary, the relaxation method requires both the sonic point
condition and the outer boundary condition, and the authors using this method
imposed the SSD properties (Keplerian rotating, radiation-dominated, etc.)
as the outer boundary condition of ADAF, i.e. the outer boundary of ADAF
had been a priori fixed to be in a state corresponding to an SSD. The
Runge-Kutta method, on the other hand, requires only one boundary condition,
and is adequate for the problem we address here. We do not know a priori
what the outer boundary condition of ADAF ought to be, so we do not specify
any; but we know for sure that black hole accretion must be transonic,
so we use the Runge-Kutta method to integrate the equations starting
from the sonic point, and observe how the solution behaves as $R$ increases.
For wrong choices of the sonic point condition and the given constant
flow parameters, the outward ADAF solution does not match an SSD solution.
Then we try again until a correct choice is made and the ADAF solution has
an asymptotic approach to an SSD solution that corresponds to the same flow
parameters as those for the ADAF solution, thus we believe that a global
solution containing an SSD-ADAF transition is found. The payment for using
the Runge-Kutta method is that not only the variables at the sonic point,
but also their derivatives there must be supplied in order to start the
integration, the calculations applying the l'H\^{o}pital rule are
troublesome, though still straightforward. (3) In Manmoto \& Kato (2000)
the transition radius $R_{tr}$ was an inputted free parameter, while in
our work it is determined by the constant flow parameters and is naturally
calculated.

Let us now comment on the relation between our results here and the 
third answer to the question of SSD-ADAF transition mentioned in 
Introduction. Takeuchi \& Mineshige (1998) suggested firstly that for 
large $\alpha \sim 1$ the thermal instability of radiation 
pressure-supported SSD could trigger the SSD-ADAF transition. 
They made time evolutionary calculations and obtained very interesting 
results: because of the dominance of radiation pressure, the SSD
becomes unstable at a radius ($\sim 3.5 R_g$ in their example solution, 
see their Figures 2 and 3), and the outer stable parts of the flow are 
disturbed and evolve towards the ADAF state; this outward propagating 
disturbance damps and stops at a large radius ($\sim 160 R_g$), then a 
transition backward to the SSD state starts from that radius, and 
propagates inward until $\sim 5R_g$ (it is larger than the original 
instability radius $\sim 3.5R_g$); finally, a two-phased flow structure 
really becomes persistent, i.e. the flow stays in the ADAF state inside 
the transition radius $R_{tr}\sim 5R_g$, and in the SSD state outside it. 
Later, Gu \& Lu (2000) made a somewhat more extensive study on the 
mechanism of thermal instability-triggered SSD-ADAF transition, 
giving an $\alpha-\dot m$ parameter diagram in which the region allowing 
the SSD-ADAF transition is clearly seen. Both Gu \& Lu (2000) and the 
present paper are for stationary flows, so if the work of Gu \& Lu (2000) 
corresponds to the first step of time evolutionary calculations of 
Takeuchi \& Mineshige (1998), i.e. it proves the cause of SSD-ADAF 
transition and shows how to determine the original instability radius where 
the transition starts to occur, then our work here corresponds to the final 
stage of Takeuchi \& Mineshige's evolutionary sequence, i.e. it 
demonstrates that a stable two-phased flow can form and exist.

In order to see more clearly the relation between the original instability 
radius $R_b$ (where the SSD solution starts to break off due to the thermal 
instability) and the transition radius $R_{tr}$ (where the SSD solution 
matches the ADAF solution in the final two-phased structure), we 
show in Figure 2 how these two radii vary with $\dot m$. In this figure the 
solid line for $R_{tr}$ is obtained by numerically solving the set of 
equations listed in $\S$ 2, with $\gamma = 1.5$, $\beta _m = 0.5$, 
$\alpha = 0.7$, and $j = 0.742 (cR_g)$ (then for each value of $\dot m$ 
a correctly chosen value of $R_s$ is required 
in order to obtain a solution that contains an SSD-ADAF transition); 
and the dashed line for $R_b$ is drawn 
by applying the instability condition $\beta \equiv p_g/(p_g+p_r) = 0.4$ 
in the standard SSD theory, which gives $R_b \propto \dot m^{16/21}$ 
(e.g. Kato, Fukue \& Mineshige 1998). The solution of Figure 1 corresponding 
to $\dot m = 0.01$ is marked by filled squares, which has
$R_{tr} \approx 12R_g$
and $R_b \approx 4.5R_g$. It is seen that $R_{tr}$ is insensitive to $\dot m$, 
and is always larger than $R_b$; and the larger $\dot m$ is, the closer 
the two radii are. It is also clear from Figure 2 that $R_{tr}$ is close to 
the central black hole, and this is because, according to 
Takeuchi \& Mineshige (1998) and Gu \& Lu (2000), the SSD-ADAF transition 
is caused by the thermal instability in the radiation pressure-supported 
region, i.e. in the very inner part of SSD. These results about $R_{tr}$ 
are distinctive from those in other transition mechanisms. For example, 
in the SSD-ADAF transition model involving radial thermal conduction, 
$R_{tr}$ appears as an inputted free parameter and has a very wide range, 
i.e. from a few to $\sim 10^4 R_g$ (Manmoto \& Kato 2000). It is worth 
studying further whether the SSD-ADAF transition radius ought to be close 
to the central black hole or it could be far away from the hole.
     
\acknowledgments

This work was supported by the National Science Foundation of China 
under Grant No.~10233030.

\clearpage

\begin{figure}
\plotone{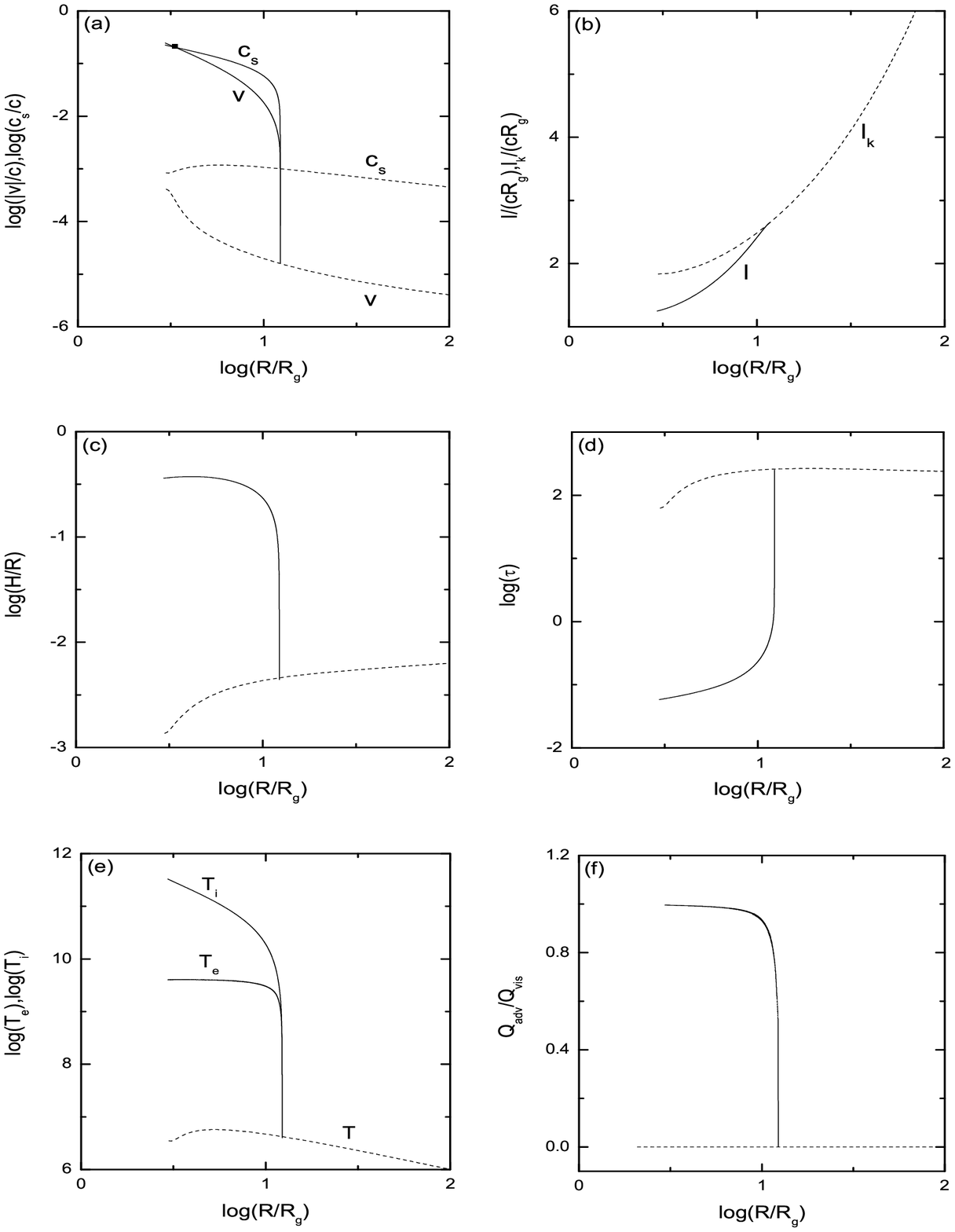}
\caption{
A global solution containing an SSD-ADAF transition. The solid line and
the dashed line represent the ADAF solution and the SSD solution, respectively.
(a), (b), (c), (d), (e), and (f) are for the radial velocity $\v$ and the 
sound speed $c_s$, the angular momentum $l$ and the Keplerian angular 
momentum $l_K$, the relative thickness $H/R$, the optical depth $\tau$, 
the ion temperature $T_i$ and the electron temperature $T_e$, and the 
advection factor $Q_{adv}/Q_{vis}$, respectively.
\label{fig1}}
\end{figure}

\clearpage

\begin{figure}
\plotone{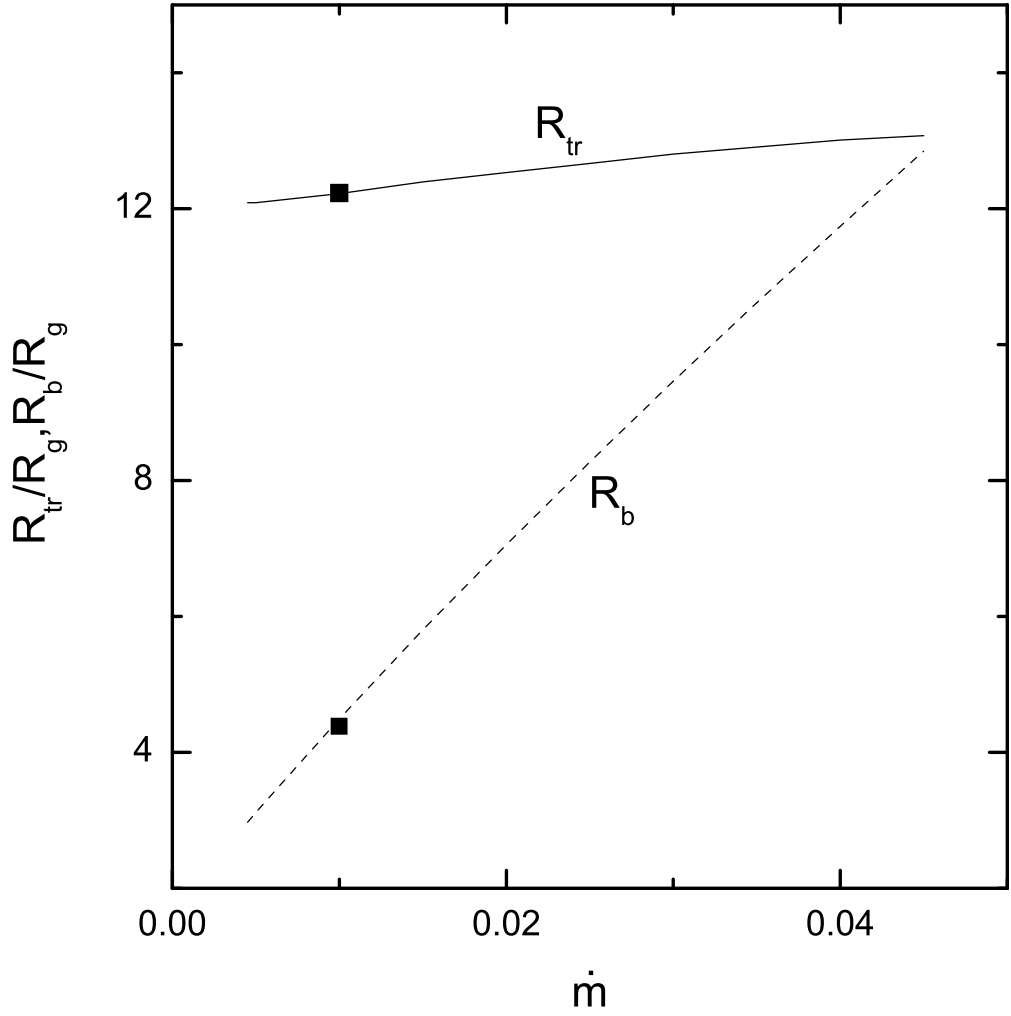}
\caption{
Dependences of the SSD-ADAF transition radius $R_{tr}$ and the thermal 
instability radius $R_{b}$ on the accretion rate $\dot m$.
}
\end{figure}

\end{document}